\documentclass[APA,Times1COL]{WileyNJDv5}

\usepackage[utf8]{inputenc}
\usepackage{subcaption}
\usepackage{natbib}
\usepackage{textcomp}
\usepackage{setspace}
\usepackage{lineno}
\articletype{Research article}% Specify the article type or omit as appropriate

\received{00}
\revised{00}
\accepted{00}

\begin{document}

%\title{Assessing the apparent sound power of birds in the field with an unattended minimal planar microphone array}
%\title{Unattended field measurement of the acoustic source level}
\title{Unattended field measurement of bird source level}

\author[1]{Guillaume Dutilleux}
\address[1]{\orgdiv{Acoustics Group, Department of Electronic Systems}, \orgname{Norwegian University of Science and Technology}, \orgaddress{7491 Trondheim}, \country{Norway}}

\corres{Corresponding author Guillaume Dutilleux \email{guillaume.dutilleux@ntnu.no}}

\authormark{Guillaume Dutilleux}
\titlemark{Apparent bird sound power}

\abstract[Abstract]{

Sound power levels or so-called source levels are essential quantities when it comes to evaluating the active space of bird species, both in the study of animal communication and when designing bioacoustic monitoring schemes. However, little data is available in the literature. In this paper I demonstrate the feasibility of the measurement of \emph{apparent} sound power in the field for bird species by using a calibrated 4-microphone horizontal array deployed on the ground. Time differences of arrival allow for the location of the sound source. Assuming a point source, the apparent sound power level is estimated after correcting for ground reflection and spherical divergence but cannot be corrected for source directivity. The benefits of my approach that is inspired from engineering measurement standards for elevated sound sources, is to minimize the contribution from ground reflections and to allow for unattended measurements, or measurements when the bird is not visible, owing to either foliage or obscurity. Moreover, my paper brings new data on sound power for 4 species of birds.   
}
%\end{abstract}

%\begin{keywords}
%\end{keywords}

\maketitle
%The audio data used to generate the results presented in this manuscript is available from \url{https://osf.io/us3ek/?view_only=b8569061acfa43cfb3e8b9cdaa852d63}. 

\section{Introduction}
A classical model of a communication system consists of three entities, first the sender, second the transmission channel and third the receiver \citep{Shannon:1949aa}. There are two obvious applications of this model in a bioacoustic context. The first one is the study of acoustic communication within a species where the concept of active space is central \citep{Naesbye-Larsen:2020aa}. The second one is the monitoring of a species with automated recorders where detection range is pivotal. By definition, within the active space the signal emitted by the sender will be successfully detected and understood by conspecifics. Within the detection range  the signal will be above the intrinsic noise level of the recording channel. In the case of bioacoustic monitoring, one implication of the detection range is the number of recorders needed to sample all the vocalizations of a given species within a specified area. Therefore, in both applications, the sound power level or the so-called source level plays a major role in the active space of bird species and in the detection range of a recorder. The sound power level is expressed in dB re $10^{-12}$ W. It quantifies the acoustic power of a source \citep{Kinsler:2000aa}. The source level is the sound pressure level in dB re $2\times 10^{-5}$ Pa measured typically at 1 m from the source along the direction of maximum emission \citep{Kinsler:2000aa}.

It is beyond the scope of this paper to review the literature on source levels in birds. The first documented field measurements I found were in the Eurasian wren \citep{Kreutzer:1974aa} and  the Common reed warbler \citep{Heuwinkel:1978aa}. A non-exhaustive list of species for which source levels have been measured is given in \cite{Naesbye-Larsen:2020aa}. At least one multi-species dedicated study of source levels and sound powers \citep{Brackenbury:1979ux} is available. This work provides field-measured source levels at 1 m and sound powers in Watts assuming hemispherical sound radiation for 17 European species of passerine birds. Spotting source level data in the literature can prove difficult because the source level is usually not the main objective of a research, but most of the time merely a by-product of this research, for instance in playback experiments where realistic sound stimuli are desirable \citep{Aubin:1998aa}. Owing to this, a new measurement of source level is not necessarily mentioned in the title or in the abstract of the relevant publications. But there is no doubt that source levels are available for only a minority of the species of vocalizing birds. In addition, when source levels are available for one species, 1) the data collected is often limited to a single site, to a few individuals, 2) the data may only correspond to a specific part of the repertoire of that species, and 3) the amount of information provided along with the measured values varies widely from just a number to a detailed measurement report.   

The approach taken in almost all these studies that provide source levels requires that the bird the source level of which is being measured is visible because the distance between the bird and the microphone must be measured, for instance with a meter stick or with an optical range finder, to be able to correct for spherical divergence. A second reason is that the position of the bird's bill must be monitored, because a bird cannot be considered as an omnidirectional sound source \citep{Brumm:2002aa} in general, although the deviation from omnidirectionality can be moderate in the few species whose directivity pattern has been investigated \citep{Witkin:1977aa, Brumm:2002aa, Patricelli:2007aa}. This prevents from studying species that are vocally active at night or species that do not vocalize from a conspicuous perch, but instead prefer to hide in the foliage. Moreover, sounds produced by small birds that vocalize while flying are probably beyond reach of this approach because measuring the distance to the source is impractical, although such measurements were made in the Rock ptarmigan, a larger species \citep{Guibard:2023aa}. Furthermore, the need for a direct measurement of distance makes it impossible to carry out unattended measurements that can be more convenient in the study of rare species or when the focus is on specific vocalizations that are known to have a low probability of occurrence. Moreover, while back-propagating to the source by only correcting for the spherical divergence is justifiable in many cases, the free field assumption is not valid in general, as the reflections from the ground and other surfaces may contribute to the received sound pressure level \citep{Embleton:1996aa}. 

But sound power is not only of interest in bioacoustics. Sound power is indeed a key characteristic in the specification of the acoustic properties of man-made products. Therefore, several basic standards have been developed in acoustical engineering for the measurement of sound power. They are based either (1) on sound pressure, or (2) on sound intensity or (3) on knowledge of a radiation factor \citep{ISO:2019aa}. It is beyond the scope of both this paper and this journal to review them here. Most of these methods require laboratory facilities. Among those that can be implemented in the field, none seems to be compatible with the measurement of the sound power of a bird, mainly because bird sounds are typically unsteady and the position of a non-captive singing bird is beyond the control of the operator. Beside these basic standards, application-specific ones are also available for outdoor sound sources. One of them specifies the measurement of the \emph{apparent} sound power of a wind turbine, where the wind turbine is modelled as an omnidirectional point source located at the hub \citep{IEC:2018aa}. Sound power is obtained from sound pressure levels that are measured on the ground. The microphones are indeed flush-mounted on rigid ground plates. The sound pressure levels are then corrected for spherical divergence and for the 6~dB amplification caused by the ground plate. With a 1/2-inch microphone the theoretical 6~dB correction holds up to the 4~kHz octave band included. This principle seems transferable to field bioacoustics, provided first that its upper frequency limit can be extended, and second that it is combined with a practical method for measuring the position of the sound source in three dimensions. 

The first issue can be addressed by turning to another ground-plate microphone configuration that was introduced to measure sound pressure levels generated by aircraft fly-bys \citep{ICAO:2017aa}. Relating to the second one, the potential applications of microphone arrays for the localization of sound sources in a terrestrial bioacoustic context had been recognized already in the late 1970s \citep{Magyar:1978aa}. Since then, many authors have used microphone arrays in various topologies \citep{Blumstein:2011aa, Rhinehart:2020aa, Verreycken:2021aa} to address a range of questions in ecology and evolution. So far, most of the research on bioacoustic localization focused on locating animals in the horizontal plane. A few authors, however, have discussed locating birds in three dimensions \citep{Stepanian:2016aa, Gayk:2020aa} or tracking their flight path \citep{Verreycken:2021aa, Dutilleux:2023aa}. To my knowledge, there is only one study that used a microphone array both to locate the source and to assess the source level \citep{Wahlberg:2003aa} in a ground-nesting bird. The location, however, was only obtained in the horizontal plane and \textit{a priori} information on source height was needed to define the source-receiver geometry completely. 

The purpose of my paper is to show that by combining a pressure-calibrated microphone with established practices borrowed from engineering acoustics for the accurate assessment of sound pressure levels and the estimation of sound power levels in the presence of ground reflections, it is possible to perform unattended field measurement of bird source level.   

%Since microphone arrays can be used to locate sound sources in 3D, the purpose of my paper is to evaluate the potential of a calibrated microphone array for the measurement of source levels and sound power levels in birds without \text{a priori} knowledge of their position. 

The remainder of this paper is organized in four steps. First, I describe the 4-microphone array on ground plates that I designed for the unattended field measurement of apparent sound power (Section \ref{sec:matmet}). This includes the procedure for retrieving the position of the bird from time differences of arrival. Second, the four study sites and the recording sessions are presented. Third, I show that the method can actually deliver measurements of apparent sound power in several bird species (Section \ref{sec:res}). %For convenience, both apparent sound power and source level are provided. 
Results for different time constants are provided. Last, I  present recommendations for further development of the method regarding flush-mounted microphones, the planar array topology and vertical accuracy, but also the format for reporting source levels, the selection of suitable site and the potential for application in other taxa (Section \ref{sec:disc}). Section \ref{sec:conclusion} concludes this paper. 

\section{Materials and methods}
\label{sec:matmet}
\subsection{Description of the measurement method}
The geometry of the method is presented in Fig. \ref{fig:TDOA_LwGeo}. The bird is represented by a point source $S$ located above the ground where microphones $M_i$ are placed. At least four pressure-calibrated audio recording channels are needed. Microphone $M_4$ is used as a reference and as the origin of the coordinate system where $z$ is the vertical coordinate. Accurate relative positioning of the microphones is required. There is no requirement for the ground to be flat nor horizontal, but the source and the microphones should be away from large reflectors like walls and buildings. 

\subsubsection{Ground-board-mounted microphones}
The microphones are neither installed on a pole nor on a conventional tripod but are ground-board-mounted instead. An inverted microphone configuration is used. This configuration is taken from standard procedures for the measurement of aircraft noise \citep{ICAO:2017aa}. It consists first of a rigid 40-cm diameter circular plate placed on the ground, and a specific mount that maintains the microphone vertically and upside down so that the gap between the microphone diaphragm and the ground board is 7~mm \citep{ICAO:2017aa}. The projection of the axis of the microphone onto the ground board is 0.15~m from the center of the plate. This configuration is equivalent to a flush-mounted microphone membrane.  

Compared to a free-field mounted microphone, the sound pressure level is 6~dB higher for a flush-mounted microphone. This correction is reported to be valid up to 10 kHz and for angles of incidence lower than 60\textdegree with the normal to the plate \citep{ICAO:2017aa}. The 6~dB increase is observed because the direct and the reflected sound have the same amplitude and the same phase \citep{Kinsler:2000aa}. In addition, the inverted microphone configuration eliminates the so-called comb filter effect in the recordings \citep{Hartmann:1997aa}. For a receiver located above ground, this interference effect is caused by the delay between the direct sound and the reflection from the ground.  Avoiding the comb filter was actually one of the motivations that led to the design of ground plates \citep{Ruijgrok:1993aa}. Furthermore, the flush-mounted configuration minimizes wind-induced noise on the microphone by comparison to a microphone on a pole or a tripod at more conventional heights because wind speed must vanish at a zero altitude relative to ground \citep{Foken:2008aa}. 

\subsubsection{Localization by time differences of arrival}
\label{sec:loc}
Assuming that a vocalization uttered by single bird from a perch has been received by all the microphones, time differences of arrival (TDOA) between the acoustic pressure signals are computed by cross-correlation. Accurate synchronization of the different recording channels is critical. The accuracy of the estimation of delays must be checked manually, for instance by looking at the alignment of spectrograms of the different channels after correcting for the delays estimated by cross-correlation. Samples where the procedure has failed must be discarded. This step leads to a first set of TDOAs identified as the \emph{measured} TDOAs in the following.  

A simple forward model of TDOA was defined where a homogeneous atmosphere at rest is assumed: 
\begin{equation}
     t_i=\frac{1}{c(T)}(d(S,M_i)-d(S,M_4))\;\forall i\in\{1,2,3\} 
\end{equation}

The model depends on 1) the unknown coordinates $(x_S, y_S, z_S)$ of the sound source $S$ representing the bird, 2) the coordinates of the microphones $(x_{Mi},y_{Mi},0)\; i\in\{1,4\}$, and 3) the speed of sound $c$ that depends on temperature.  %The TDOAs one can compute for any arbitrary source position are called \emph{modeled} TDOAs in the following. 
Optimum source coordinates are obtained by minimizing the difference between the measured TDOA and the modeled TDOA by numerical optimization \citep{Dutilleux:2023aa}. 

\begin{figure}
    \centering
    \includegraphics[width=\textwidth]{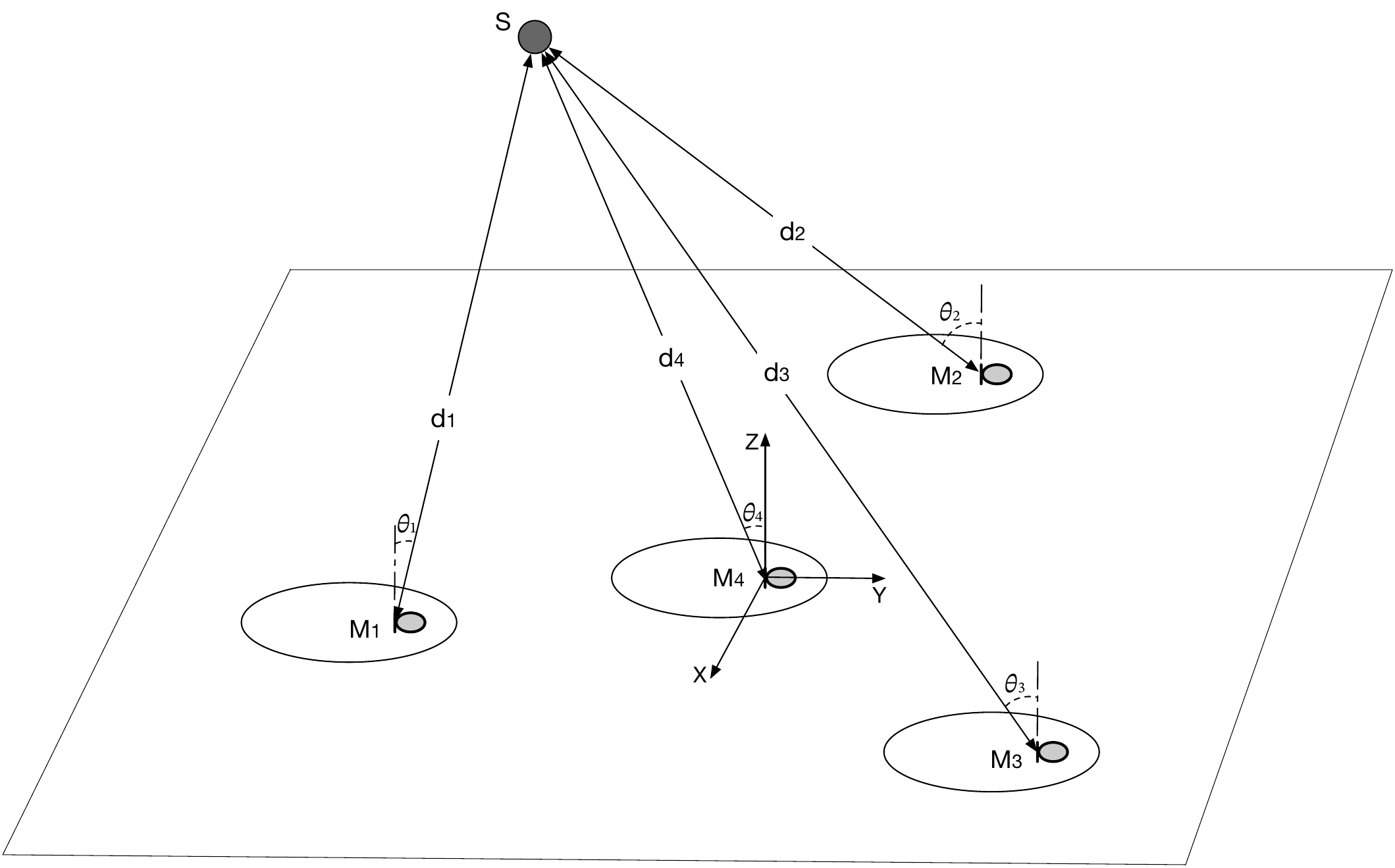}
    \caption{Geometry of the measurement of apparent sound power level, with the equivalent source $S$, the ground plates on a flat horizontal ground and the 4 microphones $M_i$.}
    \label{fig:TDOA_LwGeo}
\end{figure}

\begin{equation}
S=\underset{(x,\;y,\;z)}{\arg\min}\left(\sum_{i=1}^{i=3}(t_{i,mod}(x,\;y,\;z)-t_{i,meas})^2 + p(z)\right) 
\end{equation}

where $t_{i,mod}$ is the \emph{modelled} TDOA for microphone $i\in\{1,2,3\}$ and $t_{i,meas}$ is the \emph{measured} TDOA for microphone $i$, and

\begin{equation}
 p(z) =
  \begin{cases}
    0   & \text{if } z\geq 0\\
    z^2 &  \text{otherwise}
  \end{cases}
  \label{eq:penalty}
\end{equation}

is a penalty term to prevent the optimization algorithm from exploring negative altitudes that are mathematically acceptable because of the symmetry of the problem but non-physical.

The value of the objective function corresponding to the optimum position is compared to a threshold to evaluate the quality of the optimum found. Whenever the objective value is above the threshold, the corresponding recording is not processed further. 

Once the optimum source position $S$ attached to a labelled 4-channel audio fragment is obtained, the angle $\theta_i$ between the segment $SM_i$ and the vertical is computed and compared with the threshold value of 60\textdegree. Whenever it is exceeded, the corresponding audio track is eliminated from further calculations of sound power, since the 6 dB correction does not hold anymore in that case. The Euclidean distances $d_i$ between $S$ and $M_i$ is also computed. Assuming an equilateral array of radius 10~m around $M_4$, a numerical study of the capacity of the optimization procedure to accurately predict the correction for spherical divergence as a function of the distance, azimuth and elevation of the source in the spherical coordinate system centered on $M_4$ was carried out under the assumption of uncertain microphone positions. This study that considered 960 different hypothetical source positions showed that the accuracy of localization deteriorates as the distance to the array increases. As a consequence, when processing the audio data collected, whenever the optimized $S$ is at a distance more than twice that of the array, the corresponding recording is rejected. The stability of each optimum source position is also investigated through a Monte Carlo simulation where errors on the microphone positions (within $\pm 0.1$~m) are simulated. A Sobol sampler \citep{Sobol:1967aa} is used and 250 realizations are carried out leading to 250 optimized source positions. Whenever the standard deviation of the correction for spherical divergence between $S$ and $M_4$ is larger than $0.25$~dB the corresponding audio fragment is discarded. 

%The processing of labels included the extraction of the sound clip corresponding to each label and the same sound clip after bandpass filtering. They permitted an easy check for occasional mislabelings, for errors in the estimation of time delays and for mismatch between the bandpass filtering and the bandwidth of the vocalizations.  

\subsubsection{Estimation of acoustic quantities}
\label{sec:ac}
A calibration of each of the recording channels is required at least once per day with a reference sound source generating 94 dB re $2\times 10^{-5}$~Pa at 1 kHz. When two calibrations are performed it is advisable to check that difference in sensitivity between the two calibrations does not exceed 0.5~dB. 

For every 4-channel audio fragment that passes the tests above on the source position, and for all the microphones that meet the criterion on the angle of incidence, the exponentially time-weighted sound pressure level $L_{\tau}(t)$ is computed \citep{IEC:2013aa} in the time interval defined by the labels. The time constant $\tau=0.05$~s is used to accommodate for short contact calls, so that the same time constant could be used throughout the data collected. But for the sake of comparison with previous research, calculations according to the so-called "Fast" time-weigthing, \textit{i.e.} $\tau=0.125$~s, can be carried out as well \citep{IEC:2013aa}. The maximum sound pressure level $L_{\tau\text{max}}$ is further computed from the $L_{\tau}(t)$ over the interval defined by manual labelling. Besides the maximum sound pressure level, the equivalent sound pressure level $L_{eq}$ is also calculated. This metric is most relevant for calls that are repeated without a pause in-between or for songs.  

Like in the standardized measurement of the sound power of wind turbines \citep{IEC:2018aa}, assuming that the source is omnidirectional, the apparent sound power level $L^{'}_{W}$ based on the sound pressure level $L_{p}$ measured on the ground plate is computed as follows: 

\begin{equation}
    L^{'}_{W,i} = L_{p,i} - 6 + 10\log_{10}\frac{4\pi d_i^2}{A_0}
    \label{eq:Lp2Lw}
\end{equation},

where $i\in\{1,2,3,4\}$ is the microphone index, $L^{'}_W$ is expressed in dB re $10^{-12}$~W. On the right hand side of Eq. \ref{eq:Lp2Lw}, $-6$ corrects for the perfect reflection from the ground plate, $d_i$ is the Euclidean distance between the source and the point where the sound pressure level $L_p$ - here estimated using $L_{\text{max}}$ or $L_{eq}$ -, and $A_O=1\;m^2$ is a reference area.  

The apparent source levels $L_S$ at 1 m distance in dB re $2\times 10^{-5}$ Pa, are easily obtained from:

\begin{equation}
    L_S = L^{'}_W - 10\log_{10}(4\pi) \approx L^{'}_W - 11
    \label{eq:Lw2Ls}
\end{equation}

assuming a spherical radiation \citep{Kinsler:2000aa}. A valid source position according to the criteria listed in section \ref{sec:loc}, can provide up to 4 different values for $L_p$, because the microphone array will pick up the call from different positions. 

No frequency weighting is applied for several reasons. First the absence of weighting leads to more accurate estimates of the radiated power of a sound source. Second, depending on the context of application the receiver is not necessarily another bird. It can also be the microphone of a so-called passive acoustic recorder with a flat frequency response. Third, the standardized frequency A- and C-weightings have been developed to account for the hearing sensitivity in humans \citep{Kinsler:2000aa}. While the shape of hearing thresholds as a function of frequency is qualitatively similar in birds and humans, there are also notable differences between them \citep{Dooling:2000aa}. Therefore, I prefer to provide a 1/12-octave-band spectrum normalized to 0 dB along with the total non-frequency-weighted apparent sound power level. The exact and nominal frequencies of this fractional-octave-band frequency analysis are defined according to preferred frequencies \citep{IEC:2014aa}. The 1/12-octave spectrum is obtained by recombining the narrow bands of a Welch periodogram \citep{Welch:1967aa} based on the DFT. Based on the sampling rate of the recordings the buffer size of the DFT must be selected to guarantee a sufficient number of frequency bins in all 1/12-octave bands of interest. 

%The signal-to-noise ratio of each measurement of sound power was estimated by computing the equivalent sound pressure level over the time interval defined by the labels and the equivalent sound pressure level over the same duration but right before the interval used in the measurement. 

Compared to the existing literature on source levels, the price of unattended measurements is that while my inverse procedure provides an estimate for the source position in 3D, it does not tell anything about the azimuth of the bird's bill with respect to the microphone array. One can assume that the probability of the azimuth follows a uniform random distribution. Searching for the maximum of the $L_{\text{max}}$ in a series of measurements of the same species is the best way to estimate the apparent sound power level when the azimuth is 0\textdegree. To reduce the sensitivity to potential outliers, the 90-th percentile is computed. This percentile is also known as $L_{10}$ in the field of environmental noise \citep{Kinsler:2000aa}. Searching for the maximum apparent sound power level is most relevant when comparing unattended measurements in the case of a species for which source levels are published. But if the intention is to use the apparent sound power level to estimate the active space of a bird during a short time interval when the azimuth can be considered constant, the median - $L_{50}$ in engineering acoustics - of the $L_{\text{max}}$ was deemed a better estimator, that can account for both the variation of the motivational state and of the azimuth. When evaluating the detection range, there is indeed no reason to assume that the microphone of a passive acoustic recorder unit will stand in front of the bird's bill.  

%If the horizontal projection of the source is expected to be inside the triangle $M1M2M3$, the different microphones will see the source under very different angles. Therefore, the amplitude of the signal is likely to vary significantly from one microphone to the next because birds cannot be seen a priori as an omnidirectional point source. Microphones located behind a bird are indeed expected to capture weaker signals than those in the front, everything else being constant. For every microphone a different value for $L_w$ will then be computed. A single number is obtained from an energy average:

%\begin{equation}
%\Bar{L}_w = 10\log_{10}\left(\sum_i10^\frac{L_{W,i}}{10}\right)
%\end{equation}

\subsection{Testing the measurement method}

\subsubsection{Study sites and recording sessions}
\label{ssec:recsessions}
\begin{figure}
    \centering
    \includegraphics[width=0.7\textwidth]{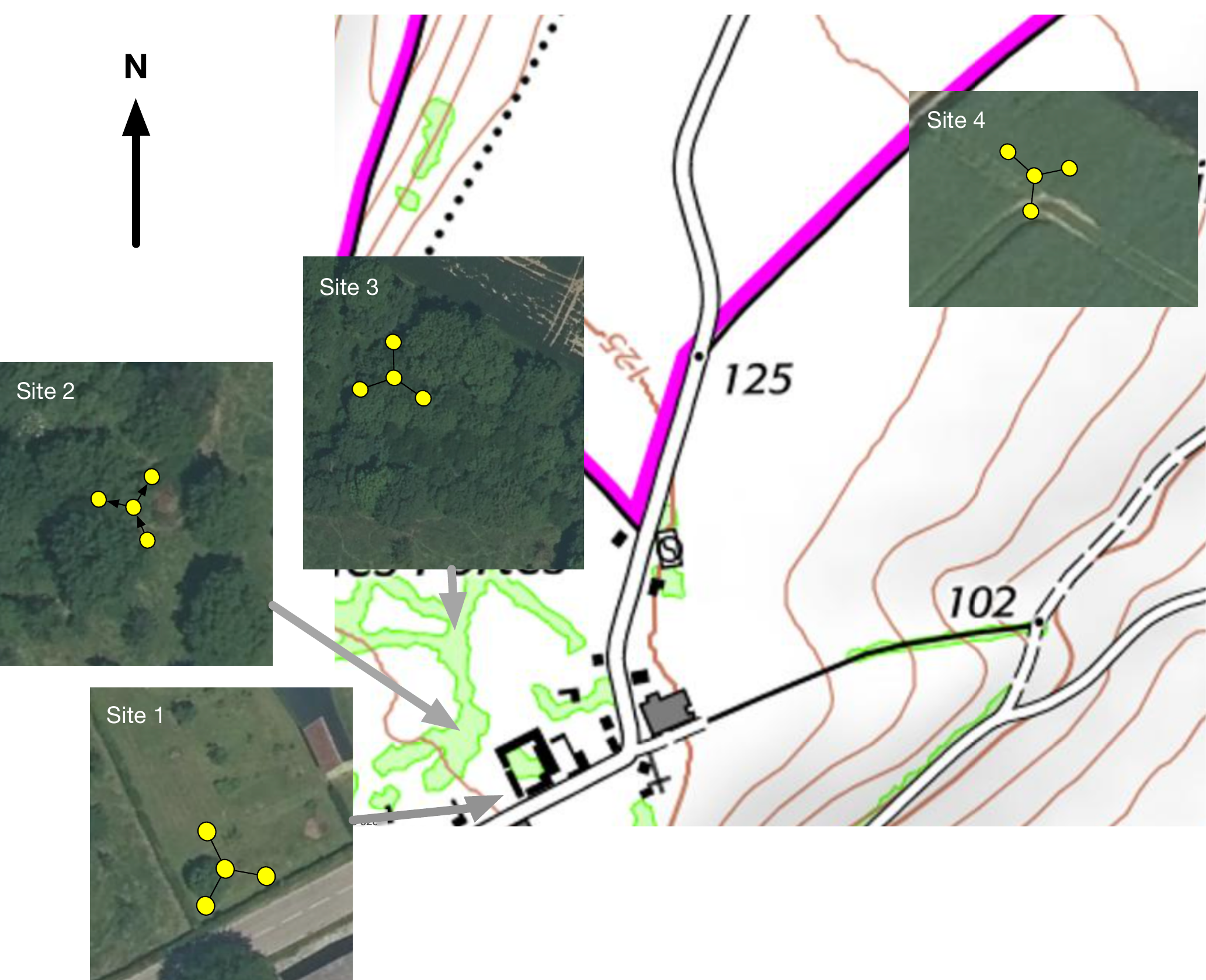}
    \caption{Overview of the 4 study sites. Site 1 is an unused kitchen garden, site 2 a pasture, site 3 a deciduous forest planted in 2009, site 4 is harvested cropland. The 3-branch stars represent the microphone arrays and their orientation. Each yellow circle stands for a microphone. Source: geoportail.fr.}
    \label{fig:sites}
\end{figure}

The method was tested during recording sessions that took place in the district of Molliens-Dreuil, Hauts-de-France, France in July 2022. In the following local summer time (UTC+2) is assumed. Four different sites were considered. At all sites the ground can be considered horizontal where the microphone array was deployed. The recordings were not attended. An overview of the study sites is presented in Fig. \ref{fig:sites}.

Site 1 was an unused kitchen garden located next to a little-trafficked road and a pasture (Figure \ref{fig:sites}). Most the area encompassed by the microphone array was covered by mowed grasses. Potential perches for birds were a wooden utility pole %(8.5~m)
, an electric power line and other wires, a young hazel tree \textit{Corylus aveilana}% (6.25~m)
, a dense hornbeam \textit{Carpinus betulus} hedge% (1.7~m)
. The closest vertical reflecting surface was an agricultural storage shed located 17~m away from the center of the array. Recordings were made from 07:52 to 12:22 on July 11th and from 05:33 to 11:27 on July 12th. 

Site 2 (Figure \ref{fig:sites}) was part of a 1.8~ha pasture planted with apple trees \textit{Malus domestica} %(7~m). 
A hornbeam \textit{Carpinus betulus} and poplars \textit{Populus sp} were also present in hedges% with heights ranging from 13.5 to 21~m
. Two houses were in the surroundings but the first reflecting surface was more than 25~m away. On this site, recordings were made from 7:55 to 10:11 on July 15th, from 21:30 on July 15th to 9:34 on July 16th. from 22:52 the same day to 9:13 on July 17th. 

Site 3 is located in a 0.5~ha deciduous forest planted in 2009 with local tree species, including Pedunculate oaks \textit{Quercus robur}, Sycamores \textit{Acer pseudoplatanus}, Norway maples \textit{Acer platanoides}, Field maples \textit{Acer campestre}, Wild cherries \textit{Prunus avium}, and European beeches \textit{Fagus sylvatica}% with an estimated average height of 14~m
. In the hedge nearby much older trees reach higher %beyond 25~m
. This site is surrounded by a pasture planted with apple trees on one side and intensively-farmed cropland on the other side (Figure \ref{fig:sites}). There, the recording session took place from July 18th at 22:00 to 10:34 the day after. 

Site 4 is in a several-ha patch of intensively-farmed cropland which explains the absence of natural perches for birds (Figure \ref{fig:sites}). At the time of recording, the patch had just been harvested, and only wheat straw remained. This apparently hostile place for biodiversity was selected because it is known by the author as a historical habitat for the Skylark \textit{Alauda arvensis}, a species that vocalizes in flight. The patch was surrounded by other crops. The distance to the same low-trafficked road as in site 1 is more than 50~m. The corresponding session took place on July 23rd from 05:35 to 07:10.

\subsubsection{Measurement equipment and data collection}
\label{ssec:recequipment}
Four pre-polarized 1/2-inch measurement microphones were used in combination with constant current preamplifiers. For the measurement points at the vertices of the triangle, a combination of a Bruel \& Kjær (B\&K, Nærum, Denmark) type 4964 microphone capsule and a B\&K 2671 preamplifier was used. For the centroid, a B\&K 4189 class I \citep{IEC:2013aa} microphone capsule was combined with a B\&K 2671 preamplifier. Use of wind screen was not necessary because there was no wind during the recording sessions.

\begin{figure}
    \centering
    \includegraphics[width=0.5\textwidth]{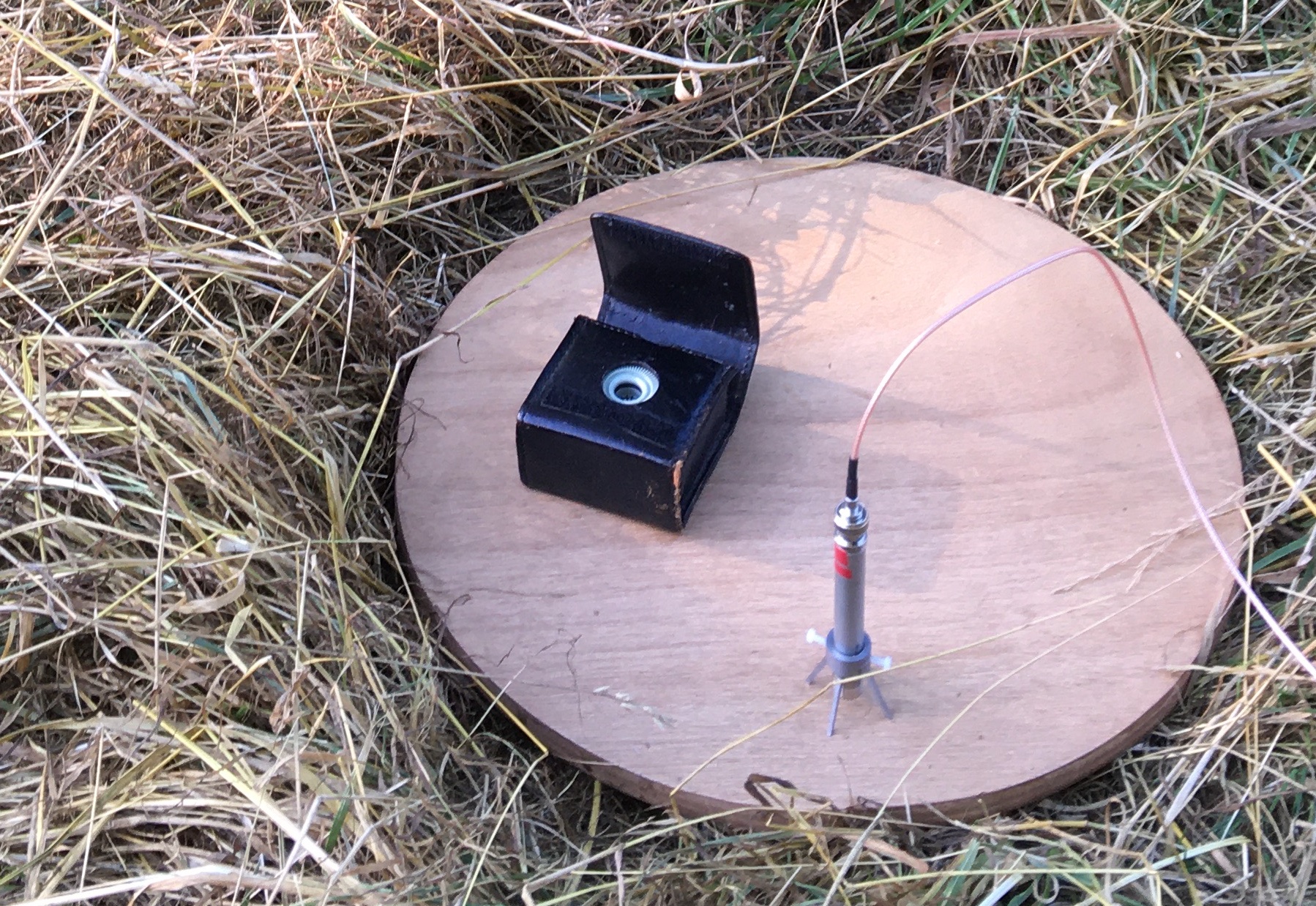}
    \caption{Close-up on the ICAO-like ground-board made of varnished plywood \citep{ICAO:2017aa}, the 1/2-inch measurement microphone in inverted configuration using a , and the calibrator used during the recording sessions.}
    \label{fig:icaoBoard}
\end{figure}

A single multichannel audio recorder (Sound Devices, Reedsburg, USA - type MixPre6) was used to collect the signals delivered by the four microphones. Coaxial cables were used to transfer both the power supply and the signals between the microphones and the recorder. Connection via cables guaranteed a sample-accurate synchronization of the signals from the different microphones. The signals were recorded in 24~bits PCM format at 44.1 kHz (site 1 and first day on site 2) or 48~kHz (other sites and site 2 second day) and stored as WAV files. Each measurement channel was calibrated at least once per day during the recording sessions with a 94~dB re $2\times 10^{-5}$~Pa 1~kHz calibrator (B\&K type 4231).

%\subsubsection{Non-acoustic parameters}
Whenever practical, horizontal distances between pairs of microphones ($\pm\;0.1$~m) were measured using a 20-m sisal rope and a 5~m tape measure. These measurements were combined with measurements of azimuth ($\pm\; 2^{\circ}$) with a hand-bearing compass (Recta, Biel, Swizerland - type KB-20/360). The horizontal coordinates of the microphones were derived from these measurements. %The heights of trees, bushes and other potential perches where measured in December 2023 with a dendrometer (Suunto, Vantaa, Finland - type PM5) at a distance of 20~m.  

Hourly temperature series for the different recording sessions were obtained from Météo-France (\texttt{meteo.data.gouv.fr}) for the Poix-de-Picardie weather station. This station was located about eight km away from the four sites. The relationship $c=20.05\sqrt{T}$ with $T$ in Kelvin was used to compute the speed of sound as a function of temperature.   
%retrieved from the Norwegian meteorology portal \texttt{yr.no}. This temperature data was checked several times by reference to an outside thermometer. Deviations did not exceed 1°C. 

\subsubsection{Annotation and post-processing}
For each recording, the 4-channel spectrogram was plotted and annotated in Audacity (Audacity Team version 3.1). Vocalizations that (1) could be attributed to a species without ambiguity, (2) were visible on each of the four tracks of the recording, and (3) were free from overlapping noise or sounds, were marked with a species-specific label spanning from the beginning to the end of the fragment of interest. Within the time interval specified by the label, simultaneous sounds in other frequency bands were not considered as overlapping sounds. The labels were set as close as possible to the beginning and to the end of each vocalization. Except for site 4, labeling focused on birds that vocalize while not flying. The span of a label was limited to a continuous vocalization. Therefore, in the case of repeated contact calls, each call was labelled individually. Identification relied here on repeated previous observations by the author that combined visual and auditory information on these very same sites. The identification of the contact calls was ascertained using on-line resources \citep{Constantine:aa}.

Further processing was fully carried out in Julia 1.9. The most notable Julia packages used are \texttt{WAV.jl}, \texttt{DSP.jl}, \texttt{Dates.jl}, \texttt{DataFrames.jl}, \texttt{OptimizationOptimJL.jl}, and \texttt{QuasiMonteCarlo.jl}. The time-weighted sound pressure level $L_{\tau}(t)$ that is introduced in section \ref{sec:ac} was implemented by the author. The implementation meets the compliance criteria on decay rate and on response to tone bursts as specified in \citep{IEC:2013aa}. 

For each label that delineates an identified vocalization, a single script will estimate the position of the equivalent sound source and, if the position meets certain criteria presented later, the script will calculate acoustic parameters for all the microphones where the angle of incidence is below 60\textdegree. Further details about the localization procedure and the estimation of acoustic quantities are given in the two sections \ref{sec:loc} and \ref{sec:ac} below. 

Each labeled vocalization was bandpass-filtered to eliminate any extraneous sounds with a 4th order zero-phase Butterworth filter \citep{Lyons:2011aa}. The cut-off frequencies were species-specific. The esimation of delays was carried out using the function \texttt{finddelay()} provided by \texttt{DSP.jl} for all species but the Short-toed treecreeper. In this species, this method is unreliable when applied on individual calls. Therefore, the cross-correlation of spectrograms along the time axis (DFT frame size 1024, overlap 1000) was used. 

While the Nelder-Mead algorithm \citep{Nelder:1965aa} was used with success in a previous study with the same array topology \citep{Dutilleux:2023aa}, here, the particle swarm optimization algorithm (PSO) \citep{Kennedy:2001aa} was preferred because the PSO belongs to the class of global optimization algorithms and the minimization task is likely multimodal. Based on a preliminary study the number of particles was set to 20.  %Simulated annealing \citep{Kirkpatrick:1953aa} was also used on the same data and returned identical opimized source positions. 
%In addition to computing the Euclidean distance between the measured and modeled TDOA, the objective function to be minimized featured a quadratic penalty term to prevent the optimization algorithm from exploring negative values of $z$. 

%For species that vocalize at low frequency, the buffer size can be set to 4096 for the computation of the DFT. For the others it is set to 1024. This guarantees a sufficient number of frequency bins in all 1/12-octave bands of interest. At higher frequencies, a smaller buffer is sufficient while making it possible to compute more spectra over the same time interval to get a better estimate of the average spectrum. A Hanning window was used \citep{Lyons:2011aa}. 

\section{Results}
\label{sec:res}

\subsection{Data collected}
The total recording time was more more than 51 hours of 4-channel audio distributed on the four sites over six days. This corresponds to more than 207 hours of single-channel audio. All the audio collected was not exploitable for different reasons. First a microphone tumbled down once on site 1 the first day. This was easily identifiable on the corresponding sound track, and caused the loss of 1h40 of recording time. On site 2 a wrong manipulation during a routine check while recording led to disarming one of the recording channels, making the final recording useless for the purpose of this study. This caused the loss of 4h09 of recording time at the end of the first night on this site. Second, stridulating invertebrates caused significant background noise on the same site in the evening and well into the night, possibly because of high air temperatures. A consequence is that localization was not possible for about 5h32 because of insufficient signal-to-noise ratio on one of the recording channels. Third, anthropogenic noise was quite present on all sites but site 4. In the first case, this was due to continuous operation of the combustion engines of large tractors and to the use of high pressure washers on the farm located near sites 1 and 2. On site 3 a harvester could be heard during the night. Due to anthropogenic noise, 5h22 of recording time had to be discarded. The remaining recording time that was amenable to annotation, source localization and apparent sound power level estimation was equal to 34h47, with 07h38 on site 1, 14h27 on site 2, 10h27 site 3 and 1h35 on site 4. 

I could identify vocalizations from 20 bird species that were present at least once on each of the four recording tracks. Several bat sounds were also picked up.
%The species that were at least once present on the four tracks of the recordings and that could be identified on sound only were the Skylark \textit{Alauda arvensis}, the Common wood pigeon \textit{Columba palumbus}, the Eurasian collared dove \textit{Streptopelia decaocto}, the European greenfinch \textit{Chloris chloris}, the Chaffinch \textit{Fringilla coelebs}, the Eurasian wren \textit{Troglodytes troglodytes}, the Blackbird \textit{Turdus merula}, the Chiffchaff \textit{Phylloscopus collybita}, the Spotted flycatcher \textit{Muscicapa striata}, the Short-toed treecreeper \textit{Certhia brachydactyla}, the Barn owl \textit{Tyto alba}, the Great tit \textit{Parus major}, the Magpie \textit{Pica pica}, the Song thrush \textit{Turdus philomelos}, a female Tawny owl \textit{Strix aluco}, the Eurasian blackcap \textit{Sylvia atricapilla}, the House sparrow \textit{Passer domesticus}. These are common species on these sites. Moreover, corvids \textit{Corvus sp.} other than the Magpie, a woodpecker and swallows appear on the recordings. Several bat sounds were also picked up. In addition, a large number of high frequency contact calls could not be identified with certainty. 
%the Barn swallow or the Western house martin 
Based on the relative occurrence of the different species in the recordings, to illustrate the operation of the method I decided to target the following six species: the Common wood pigeon, the Eurasian collared dove, the Eurasian wren, the Eurasian blackcap, the Spotted flycatcher and the Short-toed treecreeper. 
In total 606 audio fragments corresponding to one of the target species were labeled. Among them many were outside the range of 20~m from $M_4$. The remaining did not all meet the requirement of at least one angle of incidence lower than 60\textdegree. Moreover, three recordings of Eurasian wren had to be discarded because the bird was so close to $M_4$ when singing that clipping occurred, since this microphone was the only one respecting the requirement on the angle of incidence. Some vocalizations were rejected because of the sensitivity of the optimization to uncertainties on microphone positions. At the end of the selection process  141 audio fragments were deemed compatible with the assessment of apparent sound power. The results presented in the next sections are based on 109 audio fragments after eliminating 32 more fragments during the manual inspection of the estimation of delays by cross-correlation described in section \ref{sec:loc}. 
%An additional criterion was to have at least 10 calls or songs in a given species for the apparent sound power level to be computed. A consequence of this last requirement is that the data collected on site 4 was not further processed. 

\subsection{Apparent sound power}
\label{sec:aswl}
\begin{figure}
\centering
\begin{subfigure}{.5\linewidth}
    \centering
    \includegraphics[width=\linewidth]{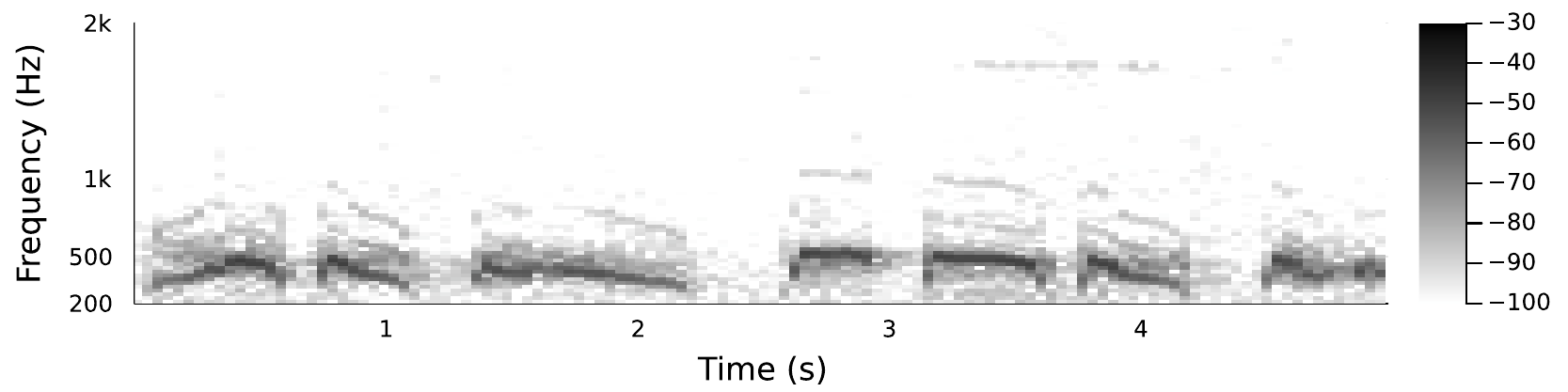}
    \caption{Common wood pigeon - spectrogram.}
\end{subfigure}
\begin{subfigure}{.45\linewidth}
    \centering
    \includegraphics[width=\linewidth]{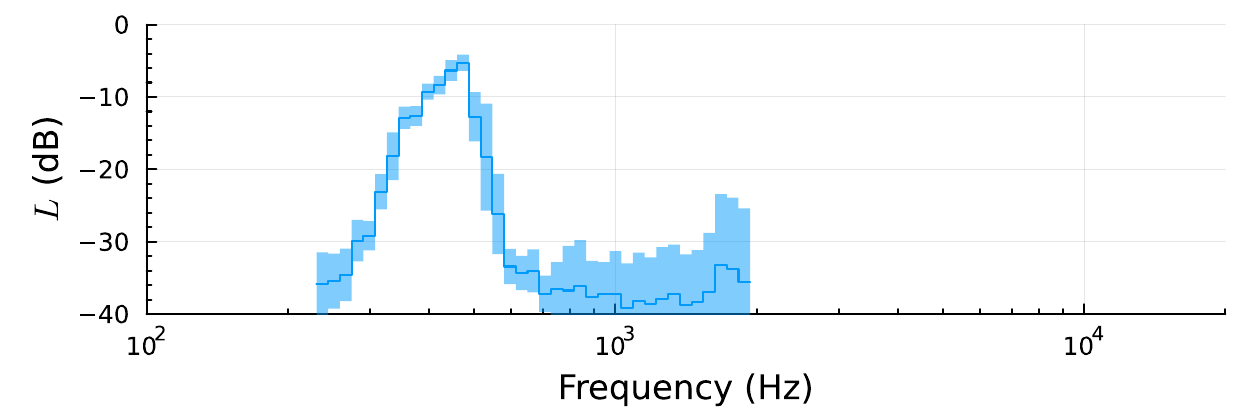}
    \caption{Common wood pigeon - normalized spectrum.}
\end{subfigure}
\hfill

\begin{subfigure}{.5\linewidth}
    \includegraphics[width=\linewidth]{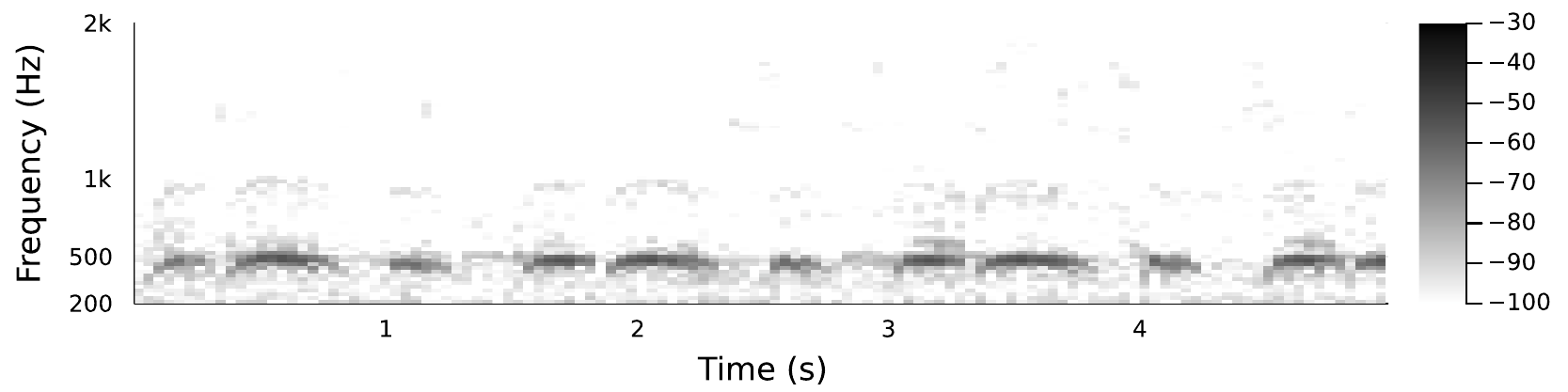}
    \caption{Eurasian collared dove - spectrogram.}    
\end{subfigure}
\begin{subfigure}{.45\linewidth}
    \includegraphics[width=\linewidth]{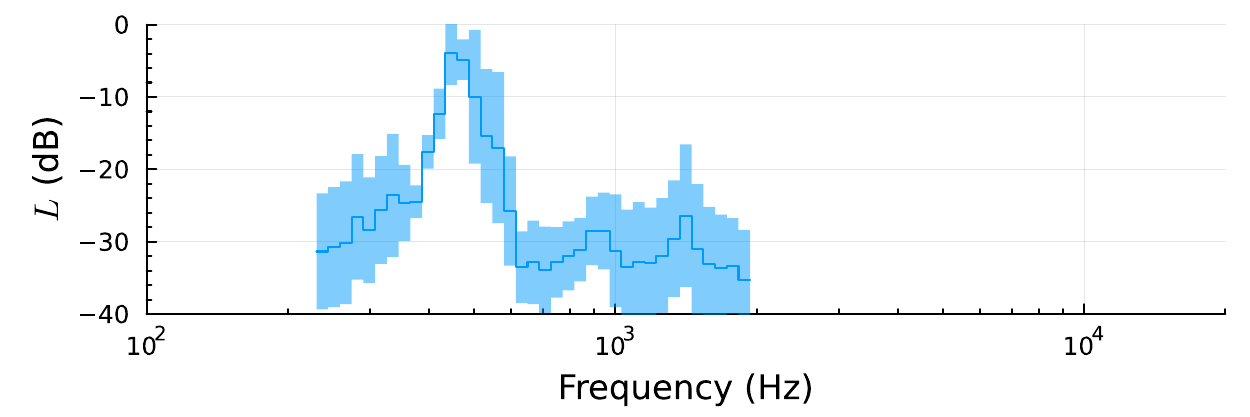}
    \caption{Eurasian collared dove - normalized spectrum.}    
\end{subfigure}
\hfill

\begin{subfigure}{.5\linewidth}
    \includegraphics[width=\linewidth]{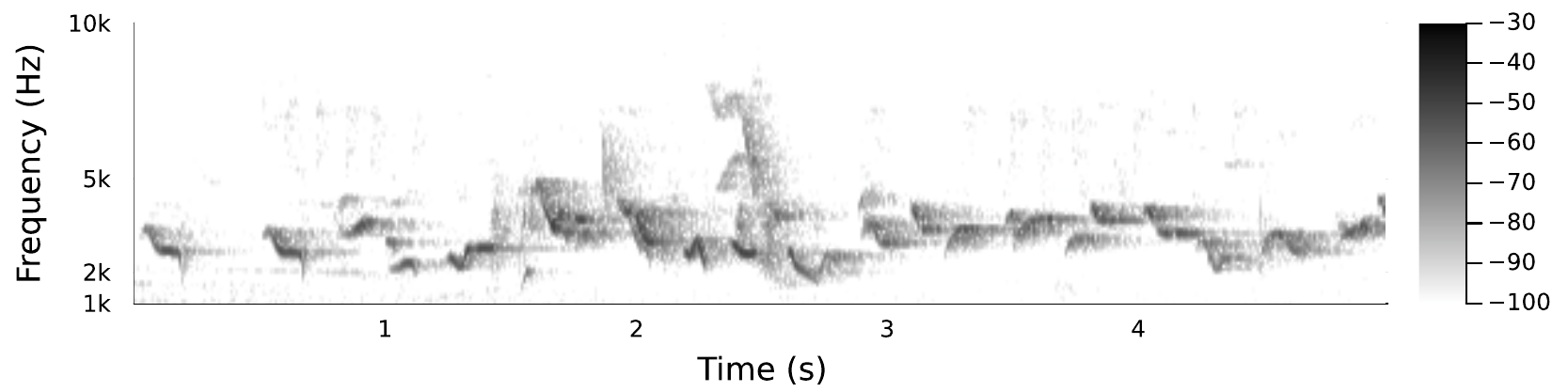}
    \caption{Eurasian blackcap - spectrogram.}
\end{subfigure}
\begin{subfigure}{.45\linewidth}
    \includegraphics[width=\linewidth]{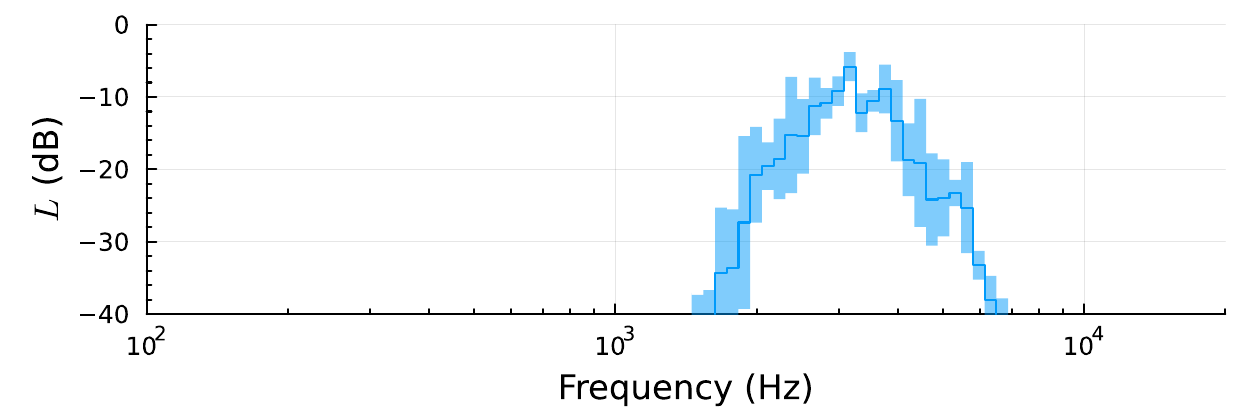}
    \caption{Eurasian blackcap - normalized spectrum.}
\end{subfigure}
\hfill

\begin{subfigure}{.5\linewidth}
    \includegraphics[width=\linewidth]{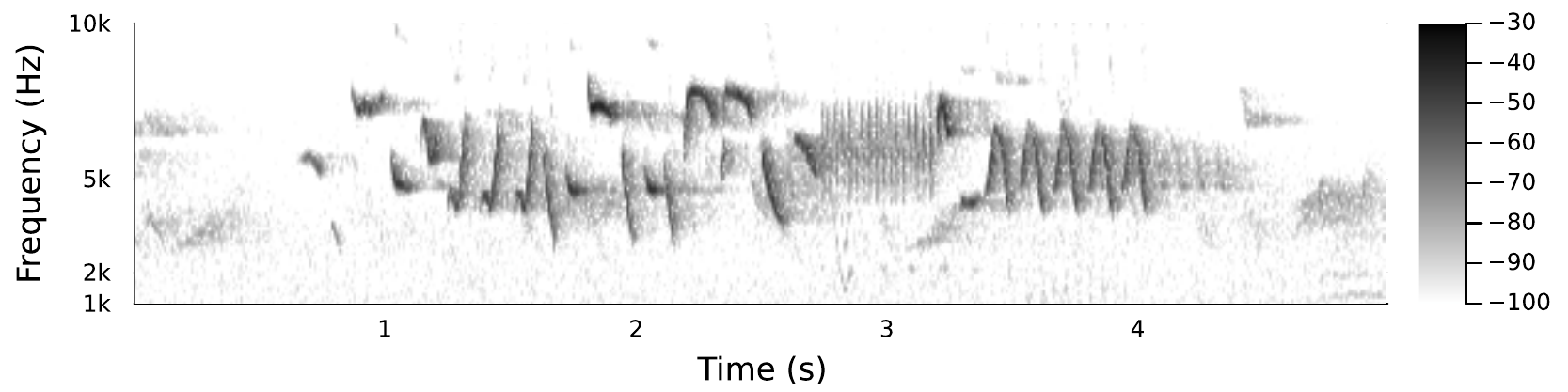}
    \caption{Eurasian wren - spectrogram.}    
\end{subfigure}
\begin{subfigure}{.45\linewidth}
    \includegraphics[width=\linewidth]{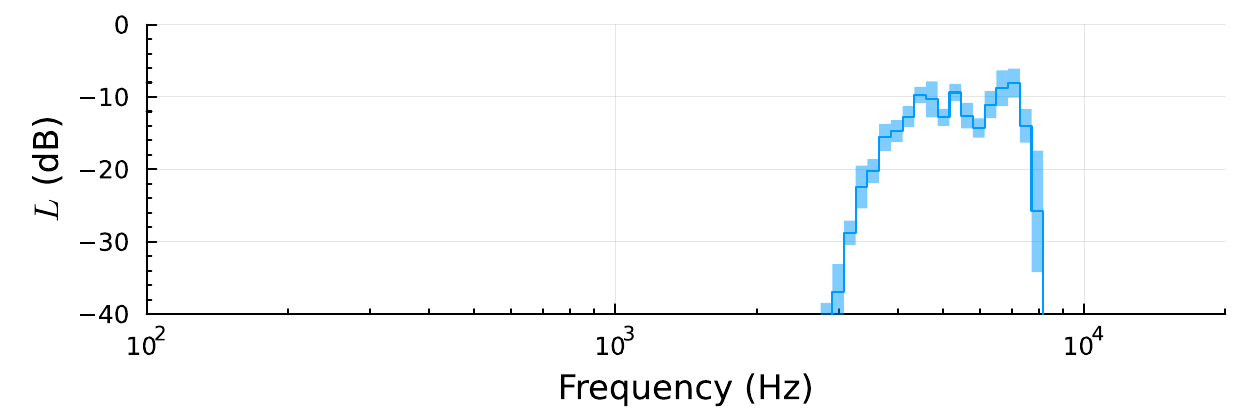}
    \caption{Eurasian wren - normalized spectrum.}    
\end{subfigure}
\hfill

\begin{subfigure}{.5\linewidth}
    \includegraphics[width=\linewidth]{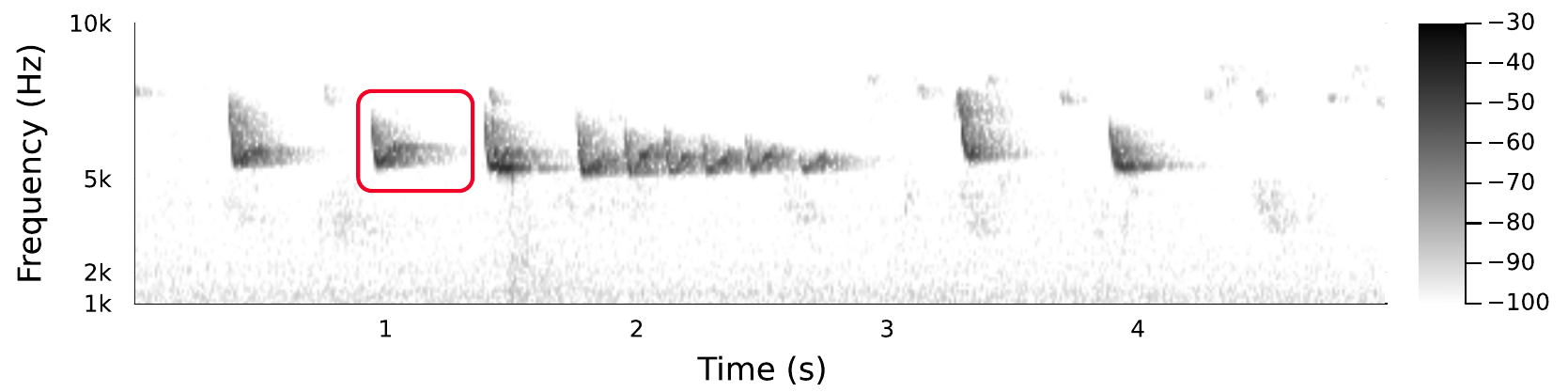}
    \caption{Short-toed treecreeper - spectrogram.}   
\end{subfigure}
\begin{subfigure}{.45\linewidth}
    \includegraphics[width=\linewidth]{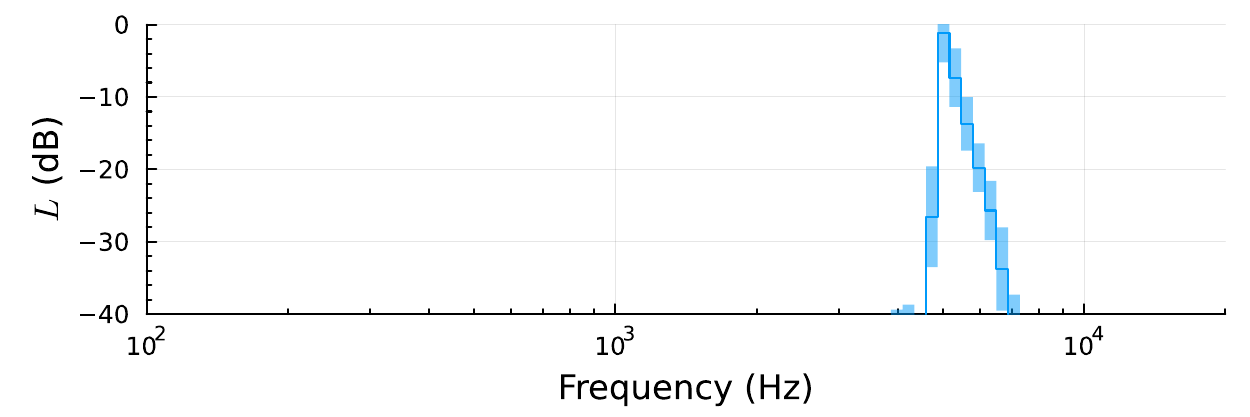}
    \caption{Short-toed treecreeper - normalized spectrum.}    
\end{subfigure}
\hfill
\begin{subfigure}{.5\linewidth}
    \includegraphics[width=\linewidth]{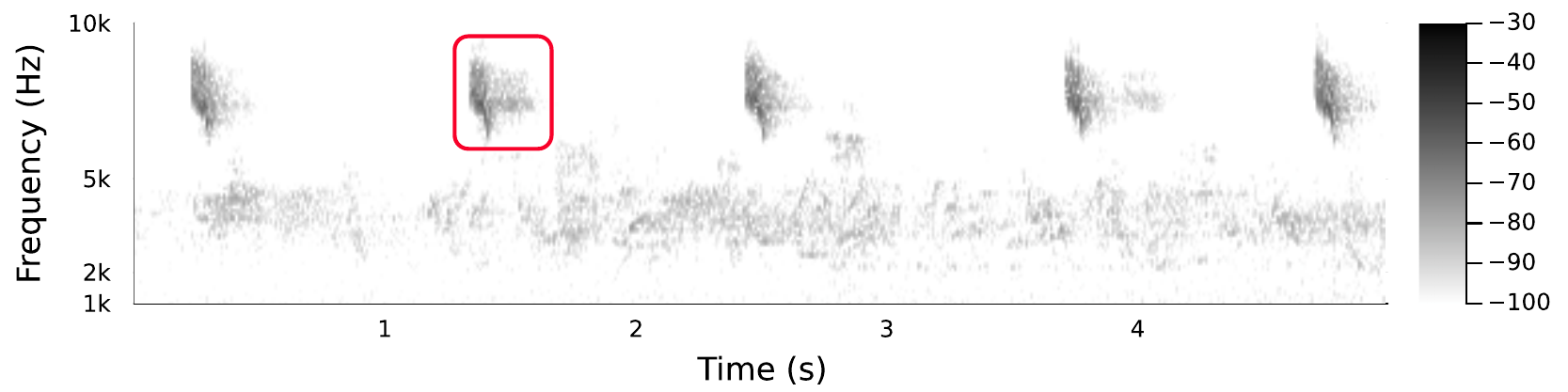}
    \caption{Spotted flycatcher - spectrogram.}
\end{subfigure}
\begin{subfigure}{.45\linewidth}
    \includegraphics[width=\linewidth]{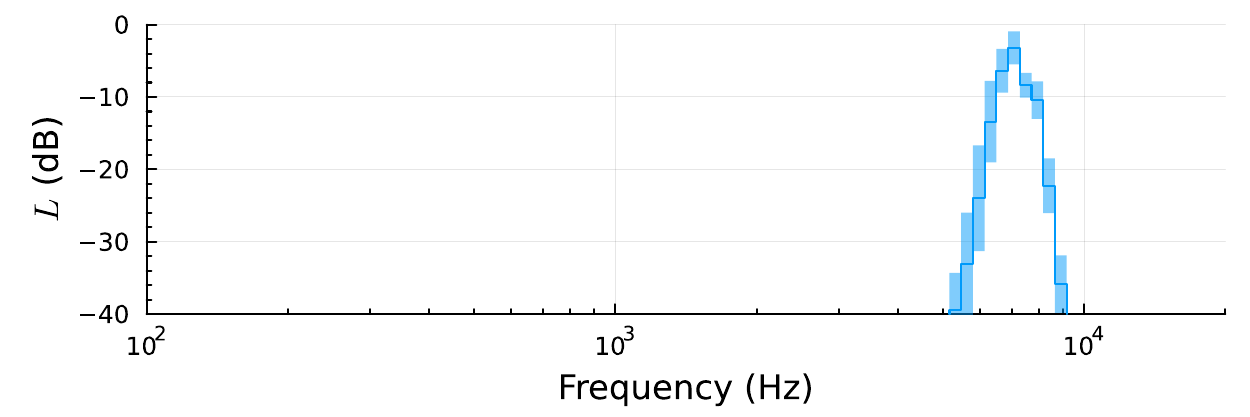}
    \caption{Spotted flycatcher - normalized spectrum.}
\end{subfigure}
\hfill

\caption{Spectral and spectro-temporal characteristics of the signals analysed. Left: typical spectrograms of the calls and songs for the six species whose apparent sound power level was computed (Table \ref{tab:ac}). On the right, average spectra in 1/12-octave bands normalized to 0 dB. The ribbon around the average corresponds to $\pm 1\sigma$. The average spectra and the standard deviations $\sigma$ were computed on 10 different non-successive sub-second calls (see the red frames in i) and k)) for the Short-toed treecreeper and the Spotted flycatcher. For the other species each averaged spectrum is based on 5 different several-second calls/songs.} \label{fig:spectra}
\end{figure}

Apparent sound power levels were computed for the six target species. %Common wood pigeon, the Eurasian wren, the Eurasian blackcap, the Eurasian collared dove, the Short-toed treecreper and the Spotted flycatcher (Table \ref{tab:ac}). 
The data used to compute the acoustic indicators is summarized in table \ref{tab:data}. For each species, figure \ref{fig:spectra} (left) provides typical spectrograms corresponding to the signals used to compute the apparent sound power level while figure \ref{fig:spectra} (right) presents an averaged spectrum for each species of interest. Among the species whose apparent sound power level was measured, the values for the Eurasian wren are based on recordings collected on 3 of the 4 sites. For three other species, the data used was collected on two sites. For the remaining species the calculations are based on a single site. The duration of the vocalizations range from about 100~ms for the Spotted flycatcher to 20 seconds for the Common wood pigeon. The vocalizations analysed include calls (Short-toed treecreeper and Spotted flycatcher), simple low frequency songs (Eurasian collared dove, Common wood pigeon) and elaborate ones at higher frequencies (Eurasian wren and blackcap). As mentioned, depending on the position of the bird with respect to the microphone array, a specific vocalization at a specific time can meet the requirement on the maximal angle of incidence up at one to four microphones. This explains the difference between the number of calls and the number of measurements in table \ref{tab:data}. The values presented in table \ref{tab:ac} can be combined with the spectra normalized to 0~dB that are presented in Fig. \ref{fig:spectra} (right) to obtain information about the apparent sound power level or the apparent source level in a given 1/12-octave band. 

\begin{table}
\centering
    \begin{tabular}{llcccc}
    \hline
     Species  & Sites & N individuals & N calls & N meas. & Tot. duration (s) \\
    \hline
         Eurasian wren & 1,2,3 & $\geq 3$ & 13 & 13 & 57 \\
     Eurasian blackcap & 2, 3 & $\geq 2$ & 15 & 39 & 96 \\ %ftn
     Common wood pigeon & 1, 2 & $\geq 2$ & 17 & 46 & 452 \\  %Cp
     Eurasian collared dove & 1, 2 & $\geq 2$ & 9 & 19 & 212 \\ %Tt 
     Spotted flycatcher & 1 & $\geq 1$ & 19 & 21 & 3 \\ %Gg
     Short-toed treecreeper & 2 & $\geq 1$ & 36 & 86 & 32 \\ %Gdj
     %\textit{Passer domesticus} & 1 & $\geq 1$ & 48 & 128 & 18 \\ %Pd
    \hline
    \end{tabular}
        \caption{Data used to compute the acoustic indicators presented in table \ref{tab:ac}.}
    \label{tab:data}
\end{table}

\begin{table}
\centering
\begin{tabular}{l|cc|ccc|c}
    \hline
     Species  & 
     \multicolumn{2}{|c|}{$L^{'}_{\text{W,50ms}\max}$} & 
     \multicolumn{3}{|c|}{$L^{'}_{\text{W,F}\max}$} & $L^{'}_{\text{Weq}}$ \\
     & $P_{50}$ & $\max$ & $P_{50}$ & $P_{90}$ & $\max$ & $P_{50}$ \\
    \hline
        Eurasian wren & 96 & 98 & 94 & 96 & 97 & 88 \\
    Eurasian blackcap & 93 & 99 & 92 & 96 & 97 & 86 \\ %ftn
     Common wood pigeon & 90 & 96 & 89 & 92 & 96 & 83 \\  %Cp
     Eurasian collared dove & 87 & 93 & 86 & 89 & 91 & 79 \\ %Tt
     %\textit{Passer domesticus} & 87 & 98 & 85 & 96 & 85 \\ %Pd
      Short-toed treecreeper & 83 & 90 & 82 & 85 & 88 &  \\ %Gdj
     Spotted flycatcher & 80 & 87 & 78 & 81 & 86 &  \\ %Gg
    
    \hline
\end{tabular}
    \caption{Acoustic indicators based on the data summarized in table \ref{tab:data}. $P_{50}$ stands for the median - or $L_{50}$, $P_{90}$ for the 90-th percentile, or $L_{10}$. The $L^{'}_{Weq}$ is not provided for the two last species because equivalent levels are most relevant for long enough signals.}
    \label{tab:ac}
\end{table}

\section{Discussion}
\label{sec:disc}
%Tm attention ce n'est pas le chant de l'aube

%Objective function valuess 
%Filtrer les hauteurs dans le grand jardin au moins. 
\subsection{Apparent sound power levels}
If one computes the source level at 1 m from the maximum of the $L_{F\max}$ using Eq. \ref{eq:Lw2Ls}, the values obtained here fall 4~dB below the published data for the Eurasian wren \citep{Kreutzer:1974aa,Brackenbury:1979ux,Camacho-Schlenker:2011aa}, and 2~dB below for the Eurasian blackcap \citep{Brackenbury:1979ux}. It would have been surprising to find higher values than in the literature, because the probability of having recorded a bird while it was singing in front of one of the microphone is low, especially when taking into account the limited number of recordings. %The time constant for the time-weighted sound pressure level is 200~ms in \citep{Brackenbury:1979ux}. This time constant was inferred from information on the equipment used \citep{Kudelski:1985aa}. As a consequence, the sound pressure level is likely slightly lower than with the standard Fast time-weighting I used, everything else being equal. Another source of uncertainty is that little information is provided on the noise metric chosen in \citep{Kreutzer:1974aa,Camacho-Schlenker:2011aa} except that it is specified that no frequency weighting was applied \citep{Camacho-Schlenker:2011aa}. In both references, the information about the measurement conditions is limited, notably the source-receiver geometry is not described. 
When considering the median apparent sound power levels and the corresponding 90-th percentile, the fast time-weighted apparent sound power levels of the Eurasian wren and the Eurasian blackcap are within 2 dB like in literature \citep{Brackenbury:1979ux}. To my knowledge, no published source levels exist for the four other species presented here. %The Common wood pigeon has an apparent sound power level that is close to that of the Eurasian wren. This may be at odds with human perception, but human hearing is at least 3~dB less sensitive \citep{ISO:2023aa} at 500~Hz (Common wood pigeon) than around 5~kHz (Eurasian wren). While the fundamental frequencies of the song of the Common wood pigeon and the Eurasian collared dove are very close, my impression is that the Common wood pigeon was louder than the Eurasian collared dove. This is in agreement with the measured values. The number of calls analyzed in the case of the Eurasian collared dove is relatively small, so that the chances of not having picked up a song when the bird's bill was facing one of the microphones is not negligible. However, due the large wavelength - about 0.7~m - of the fundamental in the Eurasian collared dove compared to the head's dimensions, the potential underestimation of the source level due to a bird not facing the microphone while singing should be limited, as $ka\ll 1$ where $k$ is the wavenumber and $a$ is the characteristic dimension of the source \citep{Kinsler:2000aa}. It is also in agreement with my experience that the calls of the Short-toed treecreeper were louder than those of the Spotted flycatcher. Nevertheless, the main objective of this paper is to present a method, more than to provide new data on apparent sound power levels or source levels. 
Overall, the values I obtained are well within the range of those reported by \cite{Brackenbury:1979ux}. 

It remains that birds are not machines. The intensity of their vocalizations is likely to vary. For instance the Lombard effect has been documented in several species \citep{Brumm:2009aa, Schuster:2012aa, Dorado-Correa:2017aa}. Moreover, the recording sessions took place during a several-week period characterized by unusually high temperatures and the complete absence of rainfall. It could have affected the vocal behaviour of the birds.  

%Although attention was paid to selecting sites that are away from large reflectors, none of the four sites presented here but site 4 comes close to an open half space. Even in an open environment, hedges and trees rows are potential reflectors. On site 3, the recorded sounds must contain a certain amount of reverberation due to the multiple scattering by tree trunks. But in either case the spectrograms were always quite sharp, and the recordings sounded quite dry, suggesting that the direct sound was dominant. This is probably due to the fact that the reflection from the ground was added in phase to the incoming sound.
%was avoided or, to be more specific, that it was added in phase to the incoming sound. 

\subsection{Limitations of the measurement method}
Since the recordings were unattended, it is difficult to assess the number of individuals whose vocalizations have been used to obtain the apparent sound power level in a given species. Considering the distances between the different sites, a reasonable assumption is that when a species was recorded on several sites, the species was represented by a different individual on each site. At least for the Eurasian wren and the Blackcap a thorough analysis of the spectrograms could have revealed distinctive features leading to the identification of different representatives of a species on a particular site. But this was beyond the scope of my study. Moreover, identification does not seem feasible when only short contact calls are recorded. As a consequence, the best way to make sure that vocalizations from several individuals are used in the estimation the apparent sound power levels is to replicate the measurement on distant sites. %Provided that one does not study species that vocalize from hidden perches, combining the method presented here with visual observations would certainly contribute to a more accurate estimations of the number of individuals.  

Assuming that the relative positions of the microphones can be measured accurately, the most critical step in the measurement I propose is the estimation of delays, because the inversion procedure leading to the source position and the correction for spherical divergence relies on them. I could not find a single method for estimating delays that worked for all species. The shape of the signals to be cross-correlated plays a major role here. Even with species-specific adjustments, the estimation of delays failed on several recordings. It was therefore essential to validate the estimation of delays by displaying the spectrograms after correcting for the estimated delays and checking for their alignment. The effort required was moderate. Alhough I did not investigate this systematically, one can expect that the higher the signal-to-noise ratio, the more reliable the estimation of delays. So in its current state, the procedure from labelled recordings to the estimation of the apparent sound power level is not fully automated. 

As the position of the vocalizing birds was not under the control of the operator, it should not come as a surprise that a very large part of the vocalizations captured during the recording sessions was not amenable to the estimation of apparent sound power levels because the distance from the source to the array was too large or because the elevation of the source with respect to the different microphones was too low. Furthermore, visual observations would have permitted to link a vocalization to a species, especially for contact calls, so that more signals could have been analyzed. Nevertheless, the campaigns presented here delivered results for six bird species. Selecting locations around trees that are known to be usual perches for birds or locations where the potential perches are enclosed by the microphone array would help maximize the proportion of useful data, except of course in forests. Situations where the probability of seeing birds choosing a perch that is close to the center of the array, combined with increasing the number of microphones on ground plates would permit to collect precious data on horizontal directivity. To my knowledge there is very little literature on this topic \citep{Brumm:2002aa,Patricelli:2007aa}.    

While the Common wood pigeon and the Eurasian blackcap produced essentially pure tones, other birds like the Eurasian wren and the Spotted flycatcher generated harmonics that extended into the ultrasonic range. Therefore, even a 48 kHz sampling rate may not be enough to capture the complete signal. However, these higher frequency components were clearly weaker than the fundamental so that they do not seem to contribute significantly to the total sound power. Moreover, the frequency limitations inherent to the experimental setup, namely the validity of the 6~dB correction above 10 kHz for the inverted microphone configuration and the frequency range of the microphones did not permit to investigate this further.

\subsection{Measurement setup}
To my knowledge, this is the first time an ICAO plate and an inverted microphone configuration were used to record the acoustic pressure in bioacoustic research. However a recent paper used a rectangular plate and a flush-mounted microphone instead \citep{Dutilleux:2023aa}. The advantage of the ICAO configuration is the smaller form factor, the lower weight and the higher frequency for which the 6~dB correction holds, which is desirable when studying passerine birds. An issue with this setup is that manufacturing a suitable microphone stand is not straightforward. The one I used was too compact to be compatible with a standard 9-cm wind screen and was unstable, although it was only a minor issue here. It is actually possible to make stands that can accommodate for a 9-cm wind screen cut in half. Nevertheless the stability of the inverted microphone position is lower than that of the flush-mounted one. With the fourpods I used, although the center of gravity of the microphone and the preamplifier in inverted position was above the top of the fourpod, the setup proved remarkably stable. Moreover, compared to the flush-mounted microphone, the ICAO configuration is likely more sensitive to wind noise because the microphone is mounted vertically. This could be compensated for by a secondary wind screen. In this study, I installed the ground plates on a flat horizontal ground. But this is not a requirement of the method. Uneven ground and gentle slopes are compatible with the use of ground plates and an inverted microphone configuration. One should however pay attention to ensure a close and uniform contact of the plates with the ground, the absence of cavity below the plate and there should be no large object blocking the line of sight between the source and the different microphones \citep{ICAO:2017aa}. 

A planar horizontal microphone array was used here. This follows from the desire to record sound pressure on the ground, so that the spectral content of the signals is preserved from the comb-filter effect induced by reflections on the ground. This makes it simple to correct for the ground effect when back-propagating to the source. Nevertheless it is not an optimal configuration if an accurate vertical localisation of the sound source is desired. %On site 1 with perches that do not exceed 8.5~m height, three optimized source heights were obviously overestimated. The error on the correction of spherical divergence was lower than 2 dB and did not generate an outlier. 
To improve the vertical resolution of the method, it is certainly possible to combine the planar microphone array with an additional microphone placed as high as possible above the ground. It may however prove difficult to reach higher than about 6~m with portable equipment. %Future research will evaluate the actual benefits of such a 3D microphone array with respect to a planar one for the field measurement of sound power. 
One could also wonder whether the array topology that was selected is the best for this task when the microphone count is fixed. As shown in Fig. \ref{fig:TDOA_LwGeo}, it consists of a triangle of microphones plus a microphone close to the center of the triangle. A square of microphones would extend over a larger area but could not resolve the position of a source on the vertical line that meets the ground where the diagonals of the square intersect. 

It was mentioned earlier that placing the microphone close to a ground plate eliminates the comb-filter effect. Therefore one can expect a reasonably flat frequency response of the plate. Due to the finite size of the plate the microphone will pick up signals caused by edge diffraction. Still, the ICAO plate is considered acceptable for the estimation of 1/3-octave band spectra \citep{ICAO:2017aa}. This configuration appears to be better suited to the estimation of the sound power spectrum of vocalizations than the classical microphone on a tripod, in addition to getting closer to anechoic conditions and providing a 6~dB amplification of the signal. 

To reduce the overall cost of the microphone array, one could imagine to keep only one measurement microphone at the center, and to use cheaper but low self noise microphones for the rest of the array. This would still allow for both source localization and for sound power measurement, even though the number of measurements per call could not exceed one, whereas it can reach four with our set-up that relies only on measurement microphones. 

In open areas, the accuracy of source localization would benefit from the use of a differential GPS system to determine the coordinates of the different microphones. The use of differential GPS would also be convenient if the array were to be deployed on uneven ground. Here I used air temperature data from an official weather station nearby, but it is preferable to measure temperatures locally, especially in forested areas.  

\subsection{Reporting of sound levels}
In this study, the apparent sound power levels were mostly computed from exponentially time-weighted sound pressure levels $L_{\tau}(t)$. This metric is better suited to the estimation of the intensity of unsteady sounds than equivalent sound pressure levels. The time constant that was primarily chosen was $\tau=50$~ms. The standardized \emph{fast} and \emph{slow} time constants were too long to estimate the sound pressure level of short contact calls like those of the Spotted flycatcher. In addition, these standard time constants were defined with different applications and human beings in mind. The 50~ms time constant was also used for longer songs, for the sake of consistency. As the time constant chosen decrease one can expect the time-weighted sound pressure level to increase. 

For each species, I chose to report the total non-frequency-weighted level in combination with a 1/12-octave spectrum normalized to 0~dB. In the literature on bird source levels, spectral information is not common. A pioneering work reported 1/3-octave bands for the Common reed warbler \citep{Heuwinkel:1978aa}. A more recent study contributed a narrowband spectrum for the Hooded crow \citep{Kragh-Jensen:2008aa}. Moreover, this is not the first time this fractional-octave-band analysis is used \citep{Frommolt:2003aa} in bioacoustics. In my opinion, the 1/3-octave band analysis is not detailed enough to describe bird vocalizations and more suited to the description of broadband anthropogenic noise. In contrast, the narrowband spectrum gives plenty of detail but is difficult to reuse by others, because it is cumbersome to report a long series of tabulated values so that the most reasonable solution is a graphical representation. Unfortunately, retrieving numerical values from a figure is not convenient either. Therefore, I preferred the 1/12-octave band analysis as a reasonable middle ground between the level of detail and the ease of reuse. For many bird species whose emission spectrum is limited to a relatively narrow frequency interval, only a small number of 1/12-octave bands is needed to document the acoustic emissions of these species. Normalized spectra are a common practice in noise engineering. Noise prediction methods rely heavily on this representation to describe the noise emissions from road or railway vehicles although in octave or third-octave bands \citep{Besnard2011,EU:2015aa}.

%Following this approach a single person can deploy, calibrate and position the necessary equipment in less than one hour.
%Orientation, directivity, borne inférieure
%Type de site: perchoir identifié, poteau, artbre isolé, plutôt que milieu ouvert. 
%Multiple birds: overlap, 
%Accès à la directivité par augmentation 
%Apparent sound power to be on the safe side but not possibility to fit a model. 

\subsection{Potential application to other taxa}
So far, the literature on source levels has mostly focused on songbirds that are very vocal like the Common reed warbler \citep{Heuwinkel:1978aa}, the Eurasian wren \citep{Brackenbury:1979ux}, or the Nightingale \citep{Brumm:2004aa} for instance. As illustrated in the results, the unattended measurement method that was outlined makes it easier to collect data for species that do not vocalize from a conspicuous perch, or for species that produce short vocalizations or for contact calls in general. 

As mentioned in the results, bat sonar signals showed up in the recordings, especially on site 2. In theory, the measurement principle described here can be extended to higher frequencies, provided that atmospheric attenuation \citep{Kinsler:2000aa} is taken into account along with spherical divergence in the back-propagation step. Moreover the setup should be deployed in open environments with little vegetation above herbaceous plants so that scattering is limited. This would also require microphones with a smaller diameter and to turn to another microphone configuration where the microphone membrane is in the same plane as the ground plate. 

Some insects, like cicadas, and some anurans, in \textit{Hylidae} for instance, are known to produce sounds from trees. In that context also one could imagine using the same method to measure the apparent sound power level. But the method is not compatible with chorusing. Recordings should not be post-processed when it is uncertain whether a single individual is signalling. This applies to birds as well. 

The method I propose here is not suited to species that emit sounds from the ground like the Great bittern \textit{Botaurus stellaris} or the Corn crake \textit{Crex crex}, because the minimum angle of incidence is 60\textdegree. A resized 4-microphone array could still be used to position the source in the horizontal plane. But other authors have already tackled this situation with a line array \citep{Wahlberg:2003aa}. If the ground is reasonably flat and homogeneous, their approach could be combined with a measurement of the equivalent source height \citep{golay2010} to obtain the position of the source in 3D. In the situation of close to grazing incidence, back-propagating from the receiver to the source must take the ground effect into account and thereby both the height of the source above the ground and the absorption properties of the ground. Here there would be no benefit in using a ground plate. A microphone located at a certain height above the ground would be preferable.

\section{Conclusion}
\label{sec:conclusion}
In this paper, I presented a new approach to the assessment of sound power or source level in birds, that relies on unique combination of processing of time differences of arrival and standardized practices in the measurement of noise from wind energy and from aircraft. This new method is based on a pressure-calibrated ground-borne microphone array with a low sensor count. The array is used first to locate the sound source in 3D. Once the location of the sound source is known, the pressure signal recorded by one of the microphones is back-propagated to the source by correcting for the spherical divergence and the contribution from the image source as well. The bird is modelled as an omnidirectional point source. The result of the procedure is the \emph{apparent} sound power level. 

Using this measurement method, the apparent sound power was obtained for 6 species of passerine birds. For each species, the total non-frequency-weighted apparent sound power level was provided along with the 1/12-octave band average spectrum normalized to 0~dB. For the four species whose source level was apparently reported for the first time, the levels should be considered as preliminary until additional estimations will give a better overview of the impact of inter-individual variations like fitness and body size, of potential geographical variations, of the context of communication, in addition to that of anthropogenic noise and the Lombard effect. Instead of a single number, a better description of the apparent sound power of a species is more likely a statistical distribution. Likewise, instead of a single number, providing the global source level along with a normalized frequency spectrum leads to a much more complete description of the acoustic emissions of bird species. With the proposed method this requires little effort. 

%The ICAO plate is likely used for the first time in a bioacoustic context. The benefits of the inverted microphone position is a broader frequency range than when a flush-mounted microphone is used. The plate offers a better control on the reflection from the ground. In addition, it leads to a stronger signal. Moreover, it is the only position where the reflection from the ground is fairly independent of frequency over a broad frequency range. This allows for a more accurate correction of propagation effects in the estimation of the apparent sound power level. 

Provided the species of interest can be identified by sound, this method allows to perform unattended measurements, or measurements when the bird is not visible, either because of low light conditions or because the it vocalizes from a hidden perch. It seems also feasible to use the method in the study of birds that vocalize while flying. 

As the deployment of the method takes little time, the method is also compatible with attended measurements. In that case, the presence of the operator offers the possibility to supplement the audio recordings with visual observations of the vocalizing bird. This will support species identification, and permit documenting the context of the vocalization. When possible, visual observations of the position of the bird with respect to the different microphones could be used to go beyond estimating the apparent sound power of an equivalent omnidirectional point source and to account for directivity.

\section*{Acknowledgments}
The field work was completed while I was on a sabbatical leave from NTNU at the Division of Terrestrial Ecology of the Norwegian Institute for Nature Research, Trondheim, Norway. 

I thank the workshop of the Department of Electronic Systems at the Faculty of Computer Science and Electrical Engineering of NTNU for manufacturing the microphone fourpods for the ground plates ; Delphine and Fernand Dutilleux for their help during the measurement of the distances between the microphones. 

\bibliography{birdLw}
\end{document}